# Low Latency CMOS Hardware Acceleration for Fully Connected Layers in Deep Neural Networks

Nick Iliev, *Member, IEEE,*, Amit Ranjan Trivedi, *Member, IEEE,*

*Abstract*—We present a novel low latency CMOS hardware accelerator for fully connected (FC) layers in deep neural networks (DNNs). The FC accelerator, FC-ACCL, is based on 128 8x8 or 16x16 processing elements (PEs) for matrix-vector multiplication, and 128 multiply-accumulate (MAC) units integrated with 128 High Bandwidth Memory (HBM) units for storing the pretrained weights. Micro-architectural details for CMOS ASIC implementations are presented and simulated performance is compared to recent hardware accelerators for DNNs for AlexNet and VGG 16. When comparing simulated processing latency for a 4096-1000 FC8 layer, our FC-ACCL is able to achieve 48.4 GOPS (with a 100 MHz clock) which improves on a recent FC8 layer accelerator quoted at 28.8 GOPS with a 150 MHz clock. We have achieved this considerable improvement by fully utilizing the HBM units for storing and reading out column-specific FClayer weights in 1 cycle with a novel colum-row-column schedule, and implementing a maximally parallel datapath for processing these weights with the corresponding MAC and PE units. When up-scaled to 128 16x16 PEs, for 16x16 tiles of weights, the design can reduce latency for the large FC6 layer by 60 % in AlexNet and by 3 % in VGG16 when compared to an alternative EIE solution which uses compression.

## I. Introduction

This research has been motivated by an important problem in real-time integrated hardware-software implementations of devices at the edge of the cloud : low-latency evaluation of fully connected (FC) layers for neural-network processing performed within the device. Example applications include deep CNN processing such as AlexNet or VGG-16, where a typical fully-connected layer has 4096 input features and 1000 output neuron activations. Another application is boundingbox object localization in an image using reinforcement learning for training and a Q-Network for inferencing with several fully-connected layers. Typical neural network hardware accelerators, such as Intel's Movidius and Google's TPU dedicate specific micro-instructions and micro-architectural processing resources for FC layer evaluation. FC layer evaluation is ususally a dense matrix-vector multiplication problem of considerable size. As an example, AlexNet has an FC8 layer with 4096 input neurons and 1000 outputs, which is similar to the FC8 layer in VGG-16. It has been shown [1] that dense FC layer evaluation is a major contributor to latency during CNN inferencing, when compared to the initial sparse convolutional layers. Therefore recent research has focused on hardware acceleration of FC layers in particular.

Fig. 1 shows such an FC layer which is the focus of our work. The evaluation of the FC layer in the figure, for one vector of input features, is formulated as a matrix-vector multiplication problem as shown in Fig. 2.

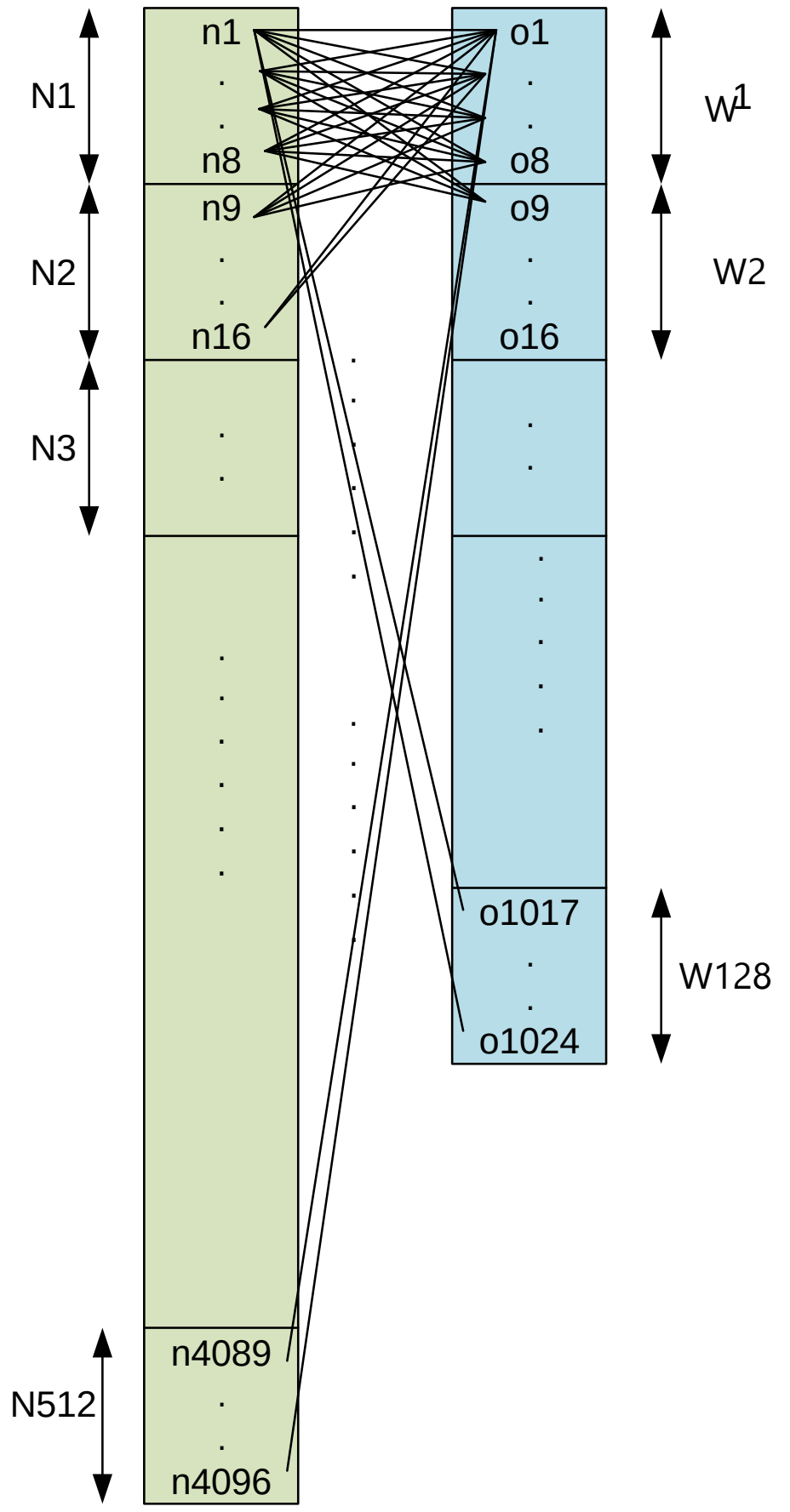

Fig. 1: Fully Connected FC8 layer in AlexNet or VGG-16 : 4096 input and 1000 outputs. Groups of 8 are indicated in the input and output vectors respectively.

Sec. II presents relation of our approach to prior works on hardware acceleration of FC layers in DNNs such as AlexNet and VGG-16. Sec. III discusses the proposed microarchitecture for our FC-ACCL design. Sec. IV presents simulation results. Sec. V concludes.

## II. Relation to Prior-Art

Hardware acceleration of DNNs has typically focused on both convolutional, CONV, and fully connected, FC, layers. This imposes some restrictions on the micro-architecture which has to handle both sparse, CONV specific kernels, as well as dense, weights based FC layers. Yuran et al. [2]

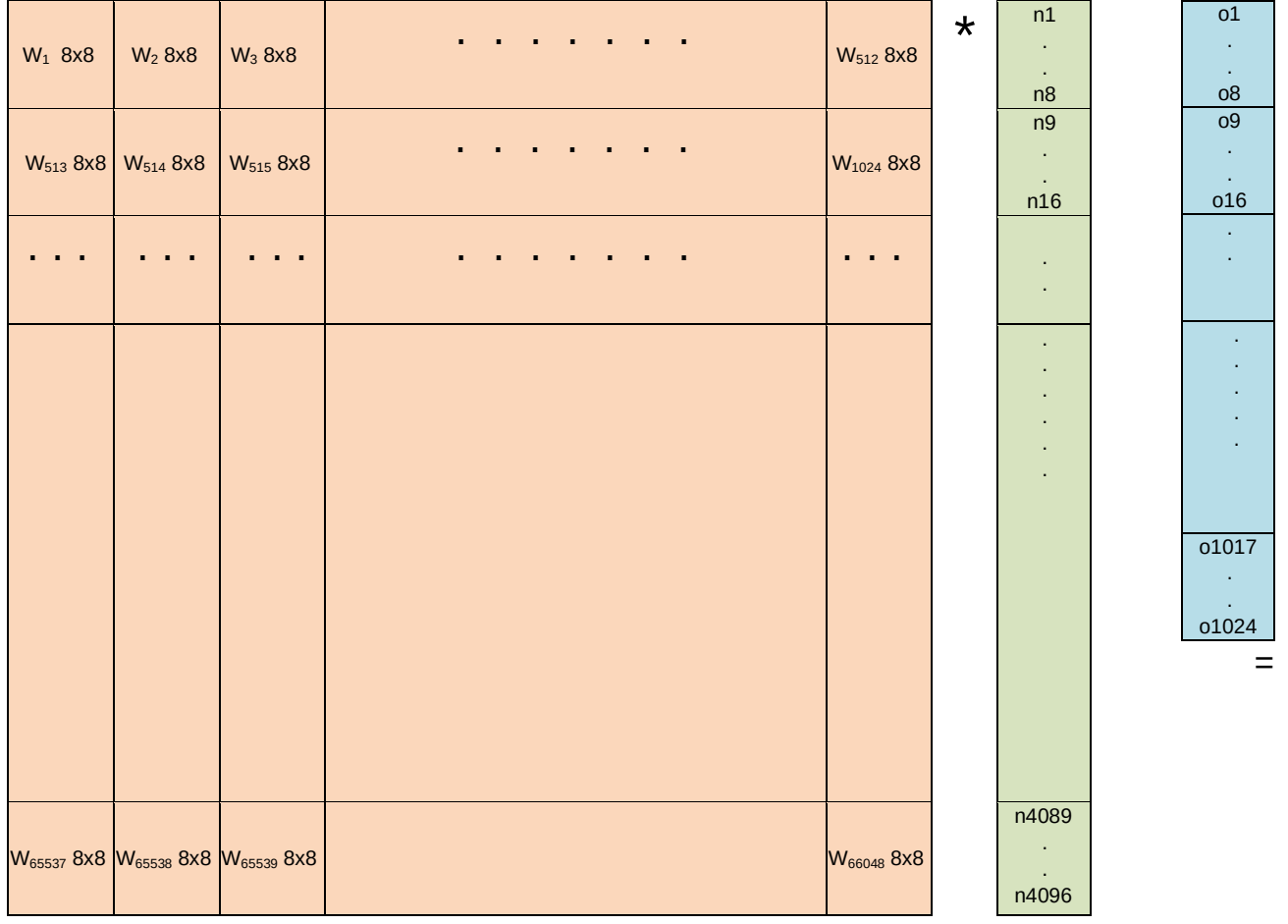

Fig. 2: The equivalent matrix-vector multiplication for the FC layer in Fig. 1. The weights are grouped in 8x8 sub-matrices (tiles) W1, W2, etc. Each column of sub-matrices is mapped, during its time-slot, to a set of 128 MACs and 128 PEs.The same set of MACs and PEs is re-used for all 512 time-slots during processing.

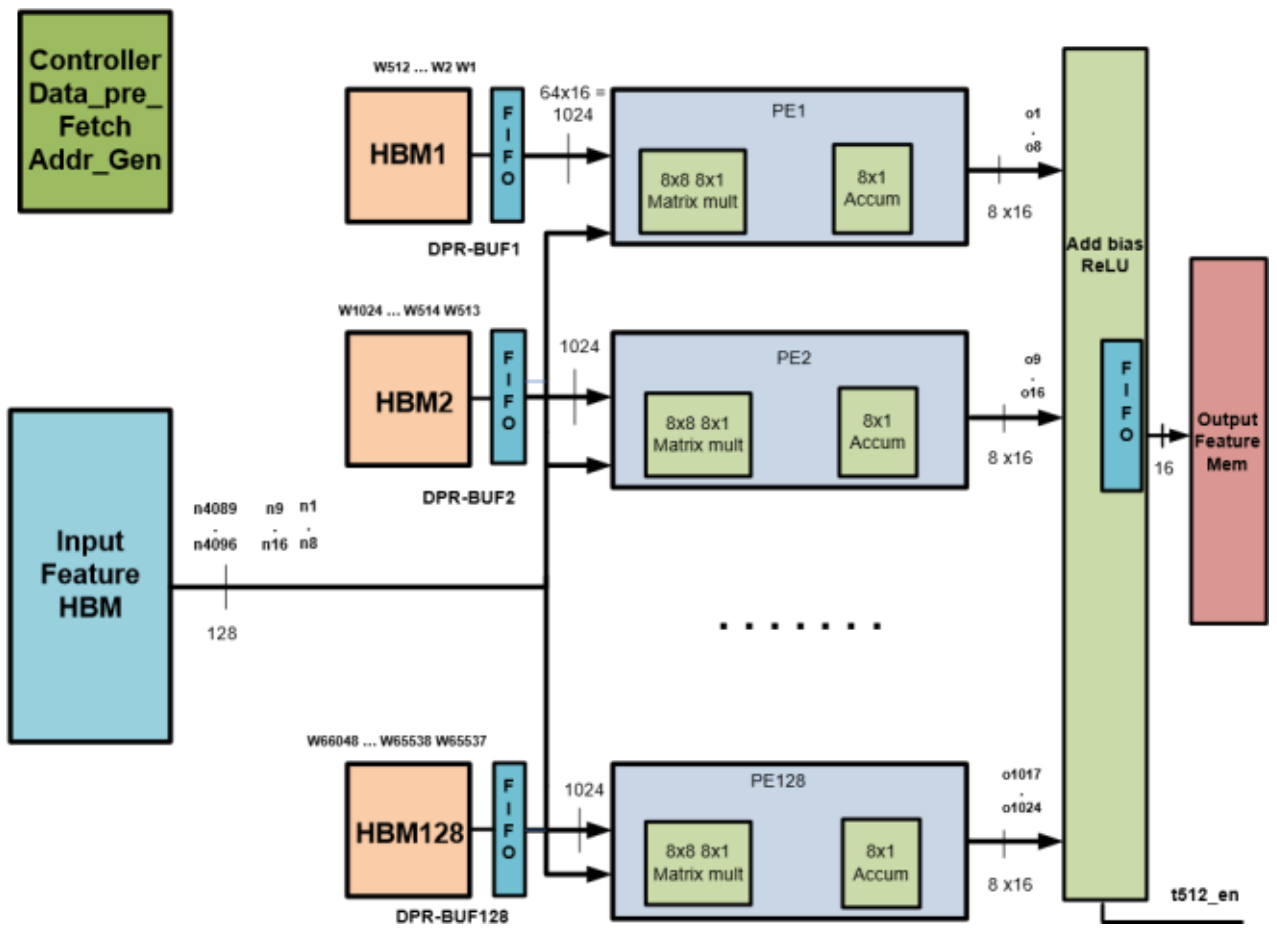

Fig. 3: High level architecture block diagram. Each HBM has dedicated datapre-fetch units and address generators. An HBM data-pre-fetch unit ensures that 1024 bits of weights are aligned for a single cycle read. Input and output memories have dedicated address generators. One top-level controller schedules the data flow in all 128 PE channels.

accelerate FC and CONV layers with a common processing element, PE, which is based on a matrix multiplier. Convolutions are unrolled to matrix multiplications for the PEs to process. The same PEs have to acccelerate the FC layers as well which can create a resource contention problem. Our solution differs from this approach since we have PEs dedicated to the FC layers only, and the sizes of the FC weights tiles (sub-matrices) are not dictated by CONV kernel and loop-unrolling considerations. Instead our PEs are optimized to reduce latency processing of the FC layer and minimize number of passes to process the entire FC layer. Jiantao et al. [3] propose to compress the FC layer weights by using Singular Value Decomposition, SVD. This approach may not always work since SVD may not exist or be numerically stable for some large FC weights matrices. In his implementation PEs are shared for CONV and FC processing and are not optimized for FC layers specifically as in our proposed FCAccel architecture. Ning et al. [4] present a global summation architecture to completely replace the matarix multiplications in the FC layers. A mathematical identity replaces multiplications with accumulators for each feature map. This places a large hardware resource requirement for FC layers with large feature maps; only small image sizes of 32x32 have been processed with the global summation method. In contrast, our FC-Accel can handle FC 25088-4096 feature layers in VGG16. Huimin et al. [5] propose an accelerator PE for both CONV and FC layers, with a batch-based computing method for the FC layers only. This differs from FC-Accel which operated on the entire FC layer ( all feature maps) and uses tiles (batches) only for the weights matrix. Their solution also has to apply two different computing patterns on FC layers which is not needed in our approach : FC-Accel uses the same computing pattern for all FC layers. Li [?] proposes a PE architecture for matrix-vector multiplication in FC layers. An entire row of weights is fetched from off-chip memory for the PEs to process. FC-Accel fetches only tiles (sub-matrices) of weights from a given column for all PEs to process and processes all rows simultaneously, column by column. The recent NVIDIA Volta GV100 architecture [6] uses Tensor Cores for matrix arithmetic. HBMs [7] are used for weights and data storage. Each Tensor Core can complete 64 floating point mixedprecision operations per clock. FC-Accel computes 128 16-bit fixed point operations per clock. The CNAPS ASIC [8] has a SIMD architecture with an array of 16x8 scalar multipliers for matrix-vector multiplication, MVM, while FC-Accel uses 8x8 or 16x16 arrays of scalar multiplier for MVM. The DianNao series of ASICs [9] implement an array of 64 16-bit integer MACs. FC-Accel uses 128 16-bit fixed-point MACs instead. The DaDianNao and ShiDianNao ASICs [10] store all weights on chip (eDRAM or SRAM) while FC-Accel uses on-chip HBMs with silicon interposers for storing weights for all FC layers and input features to these layers. Google's recently announced Edge Tensor Processing Unit, Edge TPU, [11] uses up to 65536 8-bit MAC units which limits forward inference to 8-bit precision. HBM is also used for weights and features storage. By contrast, FC-Accel maintains 16-bit fixed point precision in forward inference passes. The recently described EIE ASIC [12] accelerates both CONV and FC layers by using compression to derive a compressed network model. The resulting matrix-vector multiplications are of smaller dimentions however an 800 MHz processing clock is needed to achieve 102 GOPS for FC8 layer processing. In comparison, FC-Accel needs a 662 MHz clock for FC8 processing and achieves 1048 GOPS. The TETRIS DNN accelerator in [13] uses Hybrid Memory Cube, HMC, 3D memory which is an early form of HBM. Using 16 3D engines and 16 HMCs it can achieve 627 GOPS at 500MHz, which is 40 % lower than FC-Accel's performance.

## III. Fully Connected Layer Accelerator Architecture

We store all FC layer weights in Hight Bandwidth Memory (HBM, see JESD235A/B standard [7] ) in order to maximize the memory bandwidth of each read-out access from the weights memories. The HBM read-out bus is 1024 bits wide which allows the read out of 64 16-bit weights for each PE's matrix multiplier in 1 clock cycle. Fig. 3 shows a high-level view of the proposed architecture. It implements a column of tiles decomposition of the original weights matrix. The 128 HBMs connect to each PE's data-prefetch and on-chip buffer unit, DPR-BUF. This unit schedules a stream of two reads to two sequential column addresses so that a stream of 8 128-bit read bus cycles is generated. The following section on the data-prefetch unit details how 8 read operations from an HBM fill 8 FIFOs contained in each DPR-BUF. The 128 HBMs connect to the PEs via a silicon interposer which is not shown. The 128 PEs are reused in each of the 512 time slots which map to the 512 columns of Fig.2. The weights matrix in Fig.2 is broken up in 8x8 tiles of weights, which dictates the 8x8 PE design. Accordingly the input data is divided up into tiles of 8 elements each. Other network sizes, multiples of 8x8, are therefore possible for example 512 columns and 512 rows (square matrix in Fig.2) or 4096 inputs and 4096 outputs (FC7 in AlexNet and VGG16), 25088 inputs and 4096 outputs (FC6 in VGG16) and so on. The following sub-sections detail the micro-architecture of each PE sub-block.

Our choice of an HBM dedicated to each row of PEs avoids the need for complex 2D mesh routing and the newtorkon-chip, NoC, hardware required to implement the routing infrastructure, as required in [14] and in [13].

Notice that off-line training may produce several sets of weights (for several training optimality criteria, eg. acceptable loss function values) which can be stored in different pages in each HBM. During real time operation, between inferencing passes, a new page may be selected in some or in all HBMs and the FC layer will use a new set of weights for the next inference pass. Therefore HBM-based weights storage allows dynamic ( real-time ) weights selection between inference passes. In a following section "Up-Scaling to Larger FC Layers" we show how the proposed micro-architecture can be up-scaled for the larger FC6 and FC7 layers by using 128 16x16 PEs and their corresponding 128 HBMs.

*A. HBM Data-Prefetch Unit, DPR-BUF*

The weights tiles stored in an HBM contain a set of 64 16bit two's complement values for a specific 8x8 matrix-vector multiplier (MV-mult). The scheduler has to drive all inputs of the MV-mult in one clock cycle during its scheduled time slot. The MV-mult inputs include 8 16-bit two's complement values of input features from HBM IN (1 cycle 128 bits read out from HBM IN) as well as 64 16-bit weight values, which form a 1024 bit parallel bus of weights to the MV-mult. The DPRBUF ensures that this 1024 bit bus is driven by 8 128-bit bus outputs of each HBM as shown by 4 Da and 4 Db transactions in Fig.4. Note that an HBM's 8 DRAMs make up a stack and each DQ[127:0] output of a DRAM contributes to a portion of the DPR-BUF's 1024-bit on-chip buffer after being ratematched by its FIFO. Two clock domains, a 500MHz wr clk (write into FIFO), and a 662 MHz rd_clk (read from FIFO), are used in the DPR-BUF. This matches the HBM's 500MHz DQ[127:0] bus to the 662 MHz clock domain used in the pipeliend PEs and up to the ReLU's output FIFO write port.

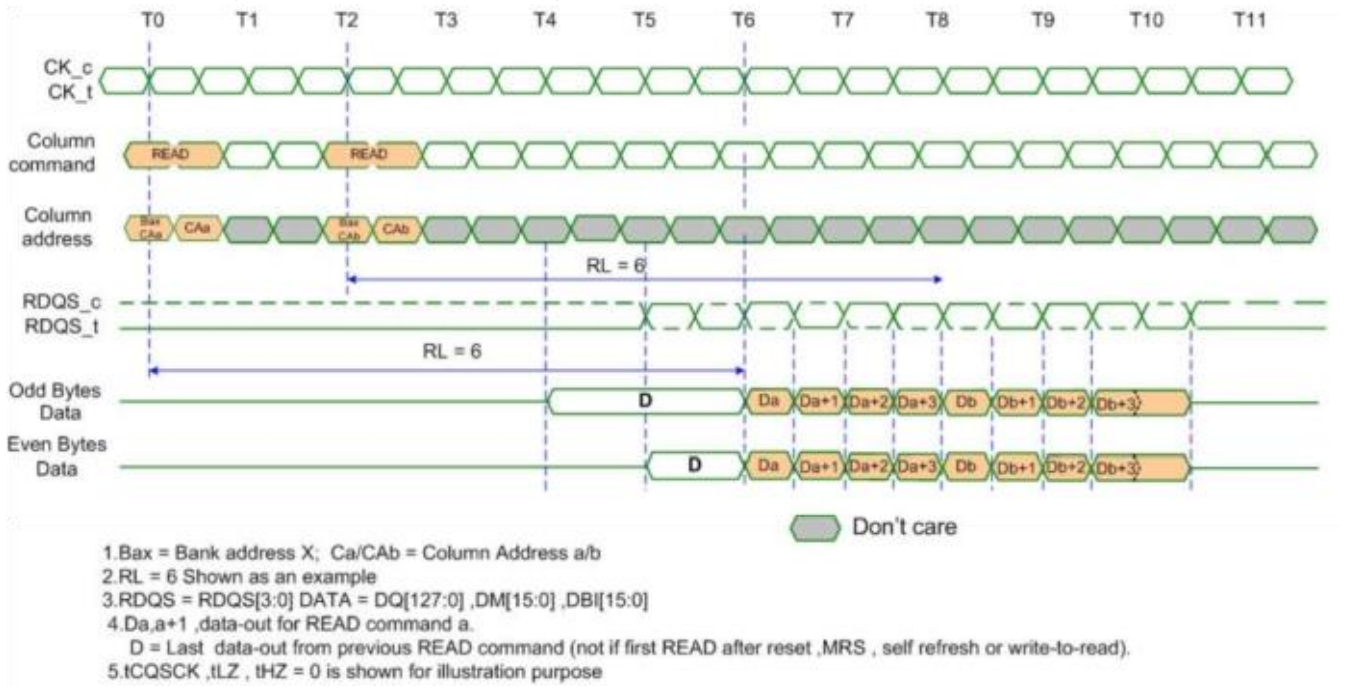

Fig. 4: Read access timing from JESD235C. The access starts with the first read request to column address Ca. After R T-cycles (T0 to T6 for R=6 example) a burst of 4 128-bit words, Da to Da+3, is available on DQ[127:0]. Similarly, the second read request to column address Cb generates a burst of 4 128-bit words, Db to Db+3. The DPR-BUF combines the 8 128-bit words and writes them into 8 corresponding FIFOs. The 8 FIFOs are then read into the 1024 bit on-chip buffer.

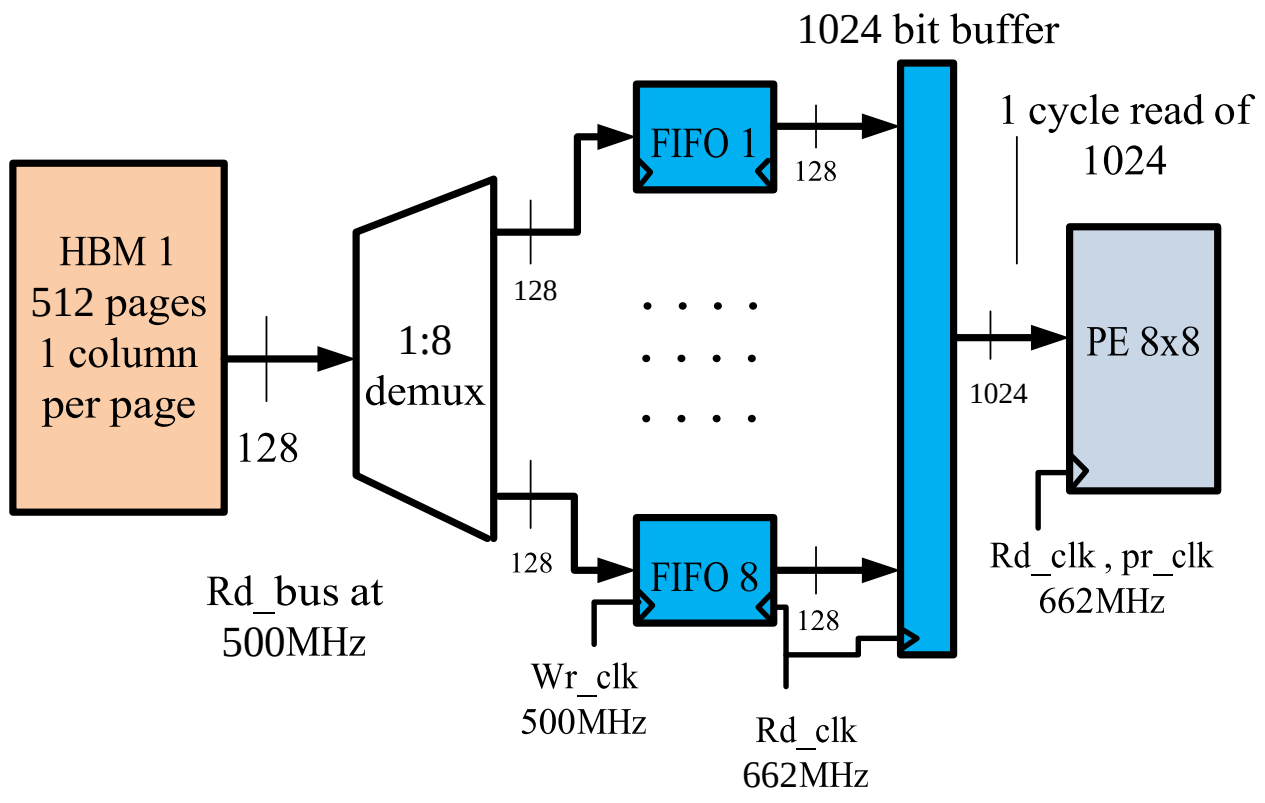

Fig. 5: Data prefetcher and on-chip buffer for HBM read accesses. One HBM is shown driving weights to its PE. The main contoller issues two read requests to column addresses Ca and Cb. Each request generates a burst of 4 128 bit transactions on DQ. The pefetcher write 8 128 bit DQ values into 8 corresponding FIFOs. A read request is then issued to all FIFOs, and their output is stored in a single 1024 bit register. This aligns the weights read-out cycle with the HBM-IN read out of the next 8 16-bit input feature values.

Fig. 4 is from the JESD235C HBM2 standard and shows how 1024 bits can be read out with two read requests, using burst length of 4, BL4, with R=6 to two column addresses in the same bank. The two read requests generate 8 128bit transactions on the DQ[127:0] bus which is sampled by the DPR-BUF. Following the main controller's sequence, the DPR-BUF initiates two read accesses to all HBMs during cycles T0 to T9 overlapped with a read access to the Input features memory HBM IN for the next input value in order for them to align at the MV-mult interface. This is shown in the following Fig. 5 .

The 8 128-bit read out cycles, in the 500 MHz clock domain, from an weights HBM (in BL4 mode) fill up its DPR-BUF's 1024-bit buffer. In the 662 MHz clock domain, the 1024 bit buffer is read in 1 cycle, Rd, overlapped with read out of the input from HBM-IN. The following 3 662 MHz cycles are processing cycles P1, P2, P3. If not empty, the FIFO is then read in the next Rd cycle and so on. In the 500 Mhz domain, the HBM is read in cycles m1 to m8. After each mx cycle, the FIFO is written in its corresponding wrx cycles. This is shown in Fig. 6. Note that cycle m8 is followed by cycle sw, to allow for HBM bank switching if the two read commands map to different banks. The main control sequence can allow for more sw cycles if needed.

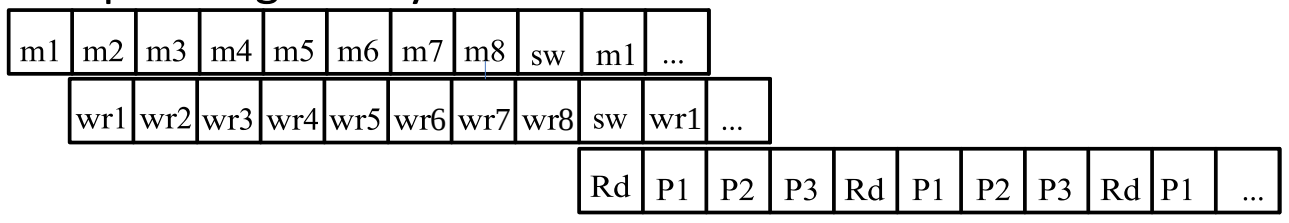

Fig. 6: DPR-BUF HBM memory read out cycles m1 to m8 in the 500MHz domain and (one of 8) FIFO write cycles wr1 to wr8. All 8 FIFOs are simultaneously read in cycle Rd and PE processing cycles P1 to P3 are in the 662 MHz domain. The 8 FIFOs are read every 4th cycle.

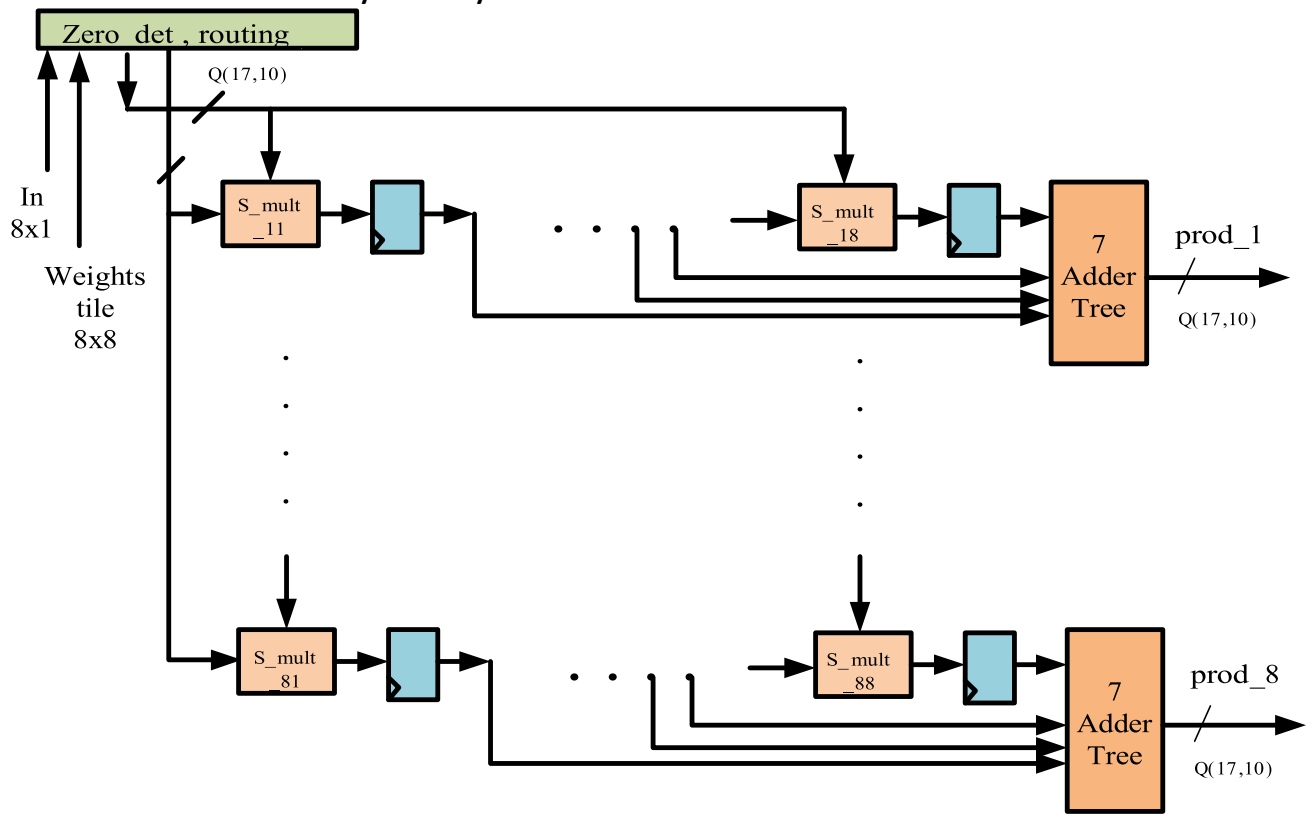

Fig. 7: MV-mult micro-architecture for 8x8 tile of weights.

*B. Matrix-Vector Multiplier Unit*

Each PE contains a dedicated 8x8 matrix-vector multiplier MV-mult for fixed-point data in the Q(17,10) format. The choice of an 8x8 tile in the weights matrix in Fig.2 determines the size of the matrix-vector multiplier as well as the number of HBMs and PEs in the system. We use 8x8 tiles of weights as an example implementation and other sizes are possible in the proposed architecture as well. MV-mult contains an array of 64 scalar multipliers where both operands have the same bit width in the Q(17,10) format. Each product is also truncated and rounded to fit into Q(17,10). The selection of 17 bits from the total of 34 bits (before truncation) is configurable and can be decided by the dynamic range of the FC layer from offline calibration. A two-stage pipeline is implemented by a dedicated register at the output of each scalar multiplier. An adder tree of seven Q(17,10) adders sums all partial products for each of the 8 rows. A zero-detector is used for each operand to gate off switching within the module when one or both operands are zero. The output 8x1 vector of products is available in 1 100 MHz clock cycle in an ASIC PDK 45 nm implementation. Fig. 7 shows the details of MV-mult. Note that for the pipelined PE ASIC implementation described later, the critical path in the seven adder tree is reduced to 1.51 nsec using a seven stage pipeline. This allowed us to run the pipelined design of a PE at 662 MHz and increase the max throughput of the accelerator considerably.

*C. Vector Accumulator Unit*

Each PE in Fig.2 has an 8x1 vector accumulation unit, VAccum, for adding up the partial products generated during each of the 512 time-slots. A V-Accum maps to each 8x1 row of the weights matrix in Fig.2; for example V-Accum-1 to o1o8, V-Accum-2 to o9-o16 and so on. Each V-Accum receives the prod-1 to prod-8 outputs from its upstream MV-mult. A new partial product is accumulated in 1 clock cycle. Fig.7 shows the details of V-Accum.

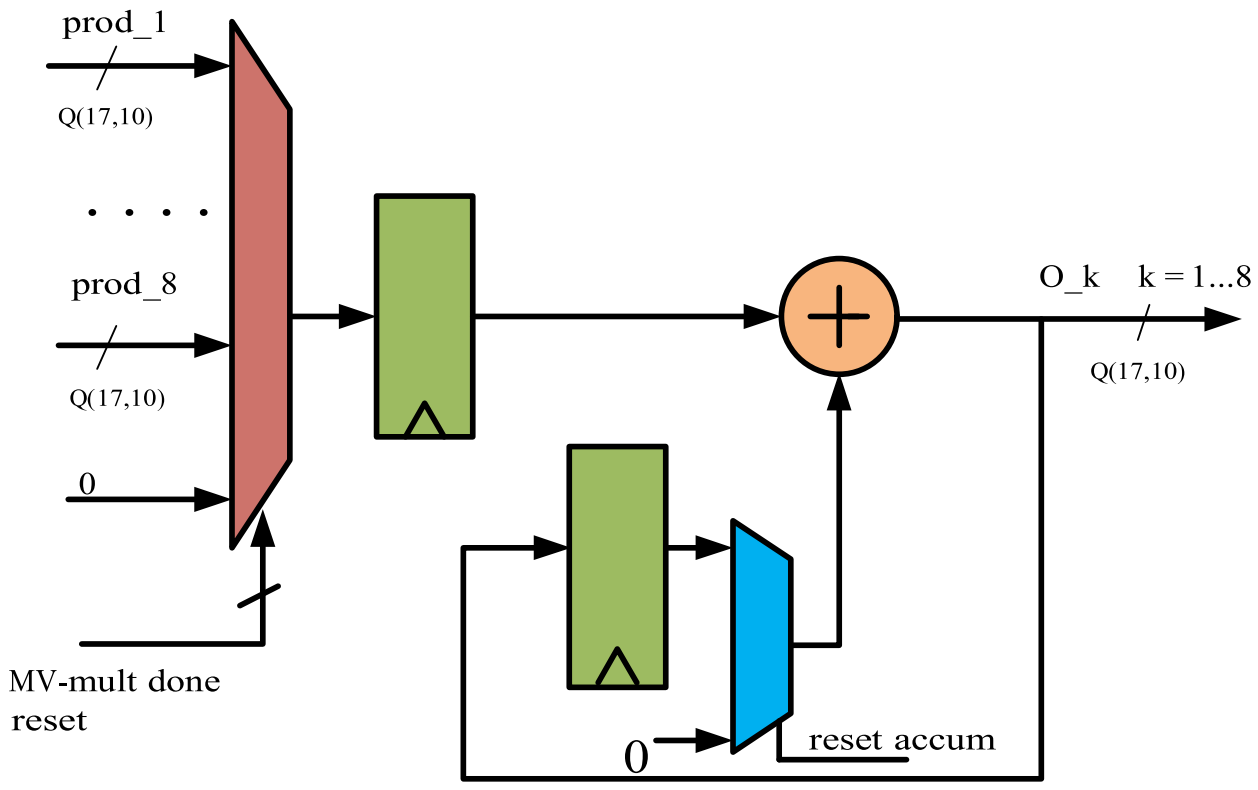

Fig. 8: Vector accumulator V-Accum datapath.

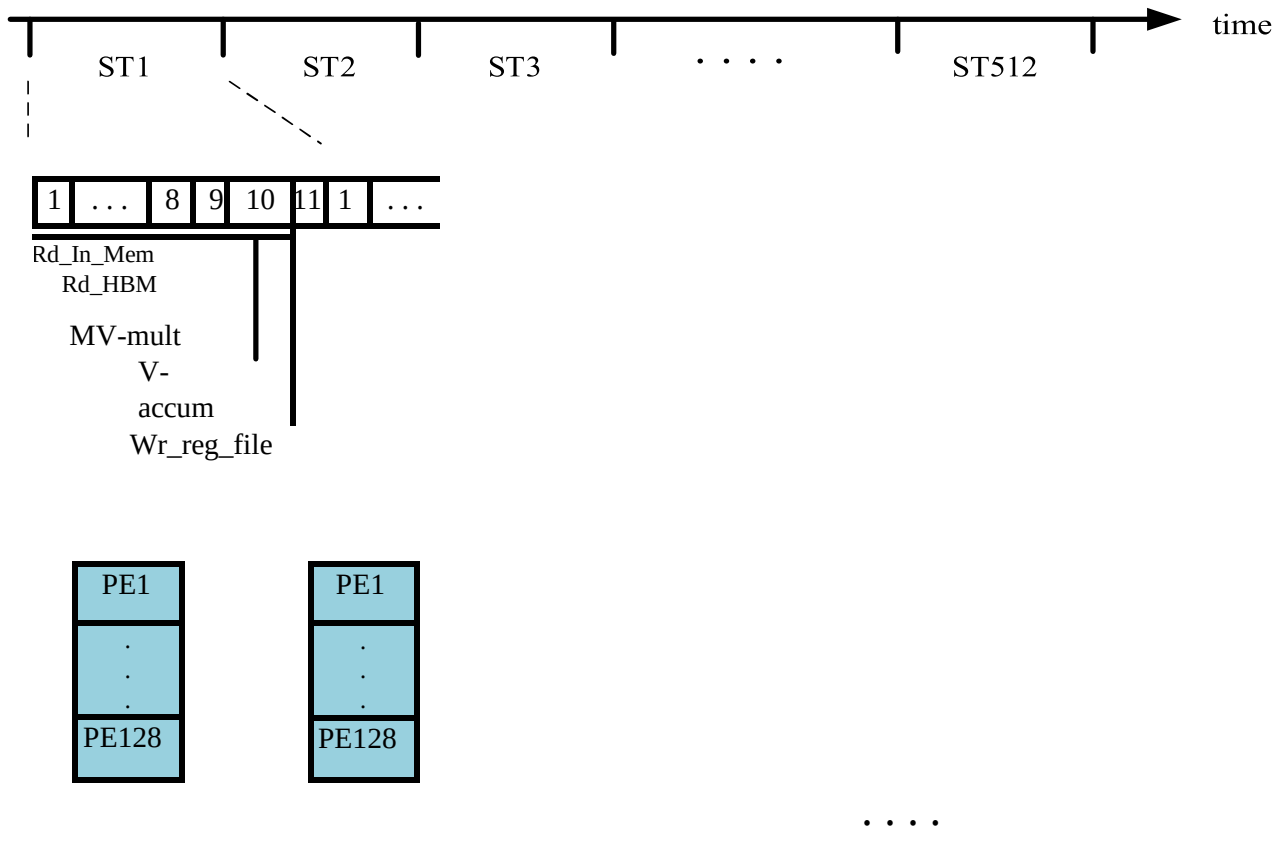

Fig. 9: Main controller sequence. All 128 PEs are processing a new 8x1 feature vector from In-MEM in each state ST1 ... ST512. In each state an 8x8 tile of weights is read from the HBM corresponding to each PE.

*D. ReLU and Bias Addition Unit*

The activation function we use is a rectified linear unit, ReLU, which introduces the max() nonlinearity as out = max( in, 0). A set of bias vectors can be added to each PE output as shown in Fig.3. Each bias vector, biasN...biasN+7, has 8 Q(17,10) elements which are added with 8 adders to the corresponding PE output vector elements oN.. oN+7. The outputs of each adder are then compared with 0. The combined addition and comparison are done in one clock cycle. Note that this is done only after the 512th (last) time-slot as indicated by the $t512_en$ signal in Fig.3. Each element is then written into a 1024 entry FIFO for streaming to the Output Feature Memory. The write clock is 100 MHz, the read clock is 150 MHz, and the FIFO implements clock-domain-crossing.

### E. Main Processing Sequence

The main controller shown in Fig.3 implements the processing sequence shown in Fig.8.

The sequence is for a 4096-1000 layer such as Alex-8 (FC8) in AlexNet or VGG-8 (FC8) in VGG16, as in Table III in [12] . Using 8x8 tiles for the weights matrix in Fig.2, the equivalent matrix of tiles is 128x512. The main control sequence therefore has 512 states, ST1 to ST512 as shown. All 128 PEs are processing an input 8x1 feature vector with their corresponding tile of weights in each state. Each column is processed in sequence, and all rows in a column are processed in parallel with each row's dedicated PE. We call this a column-row-column schedule. This schedule ensures that the computing load is equally shared among all PEs. It also achieves almost optimal load balancing among all PEs since all are utilized in each control time slot. Using this schedule we read the entire input features memory only once, where each read transaction returns a 8x1 or 16x1 vector of inputs. Similarly each row's HBM weights memory is also read only once. This minimal access pattern to the memories contributes to low processing latency and minimizes power consuption. The same control sequence, ST1 to ST512, can be applied to 4096-4096 layers such as FC7 Alex-7 and FC7 VGG-7 in [12]. To maximize throughput, 512 8x8 PEs in one pass, can be used for processing in each state. Alternatively, 128 16x16 PEs can be used in two passes, for details see section "UpScaling to Larger FC Layers" below. For the larger layers, eg.FC6 Alex-6 , 9216-4096, the control sequence has ST1 to ST1152 or 1152 states. The number of 8x8 PEs remains at 512 for one pass. For FC6 VGG-6, 25088-4096, the control sequence has ST1 to ST3136 or 3136 states.The number of 8x8 PEs remains at 512 for one pass .

## IV. SIMULATION RESULTS

### A. Simulation Setup and Comparisons to Benchmarks

The FC-Accel microarchitecture for the Alex-8/VGG-8 4096-1000 layer was implemented in fixed-point Q(17,10) ( data and weights) Verilog and simulated in ModelSim SE. A Python floating-point implementation of the same layer was used as reference. Pipelined and non-pipelined PEs were implemented with 662MHz and 100MHz clocking respectively. The seven adders tree in the original PE was pipelined to reduce it's critical path delay to 1.51 nsec and a 662MHz clock Non-zero values were used for all data features and for all weights. The following Table I summarizes the achieved processing latency for the specified design parameters and compares with recent comparable benchmarks.

TABLE I: Processing Latency Comparisons: Unit us

| Platform | AlexNet-FC8 | VGG16-FC8 |
|---|---|---|
| GPU (Titan X) Batch size 1, dense | 80.5 | 80.5 |
| GPU (Titan X) Batch size 64, dense | 5.9 | 5.9 |
| EIE with compression,pipelined PE, 800MHz | 9.9 | 8.4 |
| this work (non-pipelined 8x8 PE, 100MHz) | 56.32 | 56.32 |
| this work (pipelined 8x8 PE, 662MHz) | 8.5 | 8.5 |

The fully connected layer FC8 in both AlexNet and in VGG16 has the same 4096-1000 dimensions. Our FC-Accel latency is based on non-zero values for all input features and all weights. The data for GPU Titan X and EIE is from [12]. We summarize a pipelined 8x8 PE FC-Accel implementation using an 662 MHz clock and 7 pipeline stages for the 7 adder tree in Fig.6. The non-pipelined version uses 100 MHz clocking. Using pipelining brings the worst case critical path delay to 1.51 nsec and considerably improves latency. However it increases power dissipation as shown below.

Table II reports the operations/sec for each major processing block in FC-Accel.

TABLE II: Processing Blocks Performance

| Block | GOPS |
|---|---|
| MV-mult, non-pipelined, all 512 FC8 time slots | 1536 |
| MV-mult, pipelined, all 512 FC8 time slots | 10172 |
| V-accum , all 512 FC8 time slots | 204.8 |
| Add-bias, ReLU, final FC8 time slot | 102.4 |

The total (dynamic and leakage) power consumption in the pipelined 8x8 PE is shown in Table III for each processing block along with the cell counts.

TABLE III: Power per processing block in pipelined 8x8 PE

| Platform | AlexNet-FC8 | VGG16-FC8 |
|---|---|---|
| EIE ASIC 45nm with compression,pipelined PE, 800MHz | 102 | 102 |
| TETRIS ASIC 45nm, 500MHz, 16 3D engines, 16 HMCs | 627 | 627 |
| VC707 FPGA 28nm, 150MHz, 13.5 W | 28.8 | 131.2 |
| ZC706 FPGA 28 nm, 150MHz, 8.9 W | 16.5 | 71.2 |
| this work, 128 non-pipelined 8x8 PEs, 100MHz, 17 W | 108 | 108 |
| this work, 128 pipelined 8x8 PEs, 662MHz, 90.1 W | 1048 | 1048 |

| Block | Power | Cells |
|---|---|---|
| MV-mult 8x8, pipelined | 581.6 mW | 140662 |
| V-accum 8x1 | 12.3 mW | 2468 |
| Total PE | 593.9 mW | 143130 |

Table IV compares the achieved operations/sec for the 40961000 FC layer with other comparable benchmarks (all units are in GOPS) and shows the speedups achieved by FC-Accel in ASIC PDK 45 nm technology. The FPGA implementations are from Table 17 in [15] . The TETRIS implementation is from [13].

TABLE IV: Comparison with ASIC and FPGA Platforms for FC8 Acceleration

We note that the TETRIS accelerator uses Hybrid Memory Cube, HMC, 3D memory, an early form of HBM memory.

### B. CMOS ASIC Implementation

We have implemented the Alex-8/VGG-8 4096-1000 using the CMOS ASIC PDK 45 nm standard cell library for synthesis. The Cadence RTL Compiler (RC) tool was used and the design achieved timing closure with a 100 MHz clock for the non-pipeliend PE and 662 MHz for the pipelined PE.

Table V summarizes timing, area, and power for the non-pipelined and pipelined FC-Accel design, with 128 8x8 pipelined or non-pipelined PEs.

TABLE V: FC-Accel PDK45 Standard cell Implementation

| Design Technology | NCSU PDK 45 nm | |
|---|---|---|
| clk freq | 100 MHz non-pipelined | 662 MHz pipelined |
| std cell VDD | 1 V | 1V |
| Combinatorial gates | 11188035 | 13245537 |
| Sequential Cells (DFFs) | 313480 | 844936 |
| Dynamic power | 16.9 W | 89.8 W |
| Total power (Leakage,Dynamic) | 17.2 W | 90.1 W |

*C. Energy Efficiency Characterization*

In this section we present plots of GOPS/W at 1V PDK 45 nm for both non-pipelined and pipelined PE implementations of the proposed micro-architecture. The power estimates are for the synthesized netlist from the Cadence RC tool and assume worst case statistical switching. Power consumption due to the HBM and input features memory interfaces is not included at this time.

*D. Up-Scaling to Larger FC Layers*

In this section we present estimated performance with an upscaling of the proposed micro-architecture for the larger FC6 and FC7 layers in AlexNet and VGG16. We use a 16x16 PE for these layers in order to efficiently process the larger sizes of the weight matrices. The 16x16 tile of weights reduces the number of rows and columns of matrix in Fig.2 and simplifies the processing schedule as well. Both layers have 256 matrix rows with 16x16 PEs. Since a 16x16 PE has 256 weights for its matrix-vector multiplier, this is 256 x 16 = 4096 bits of weights for each PE in each row. The HBM for that row can be read in 4 cycles to deliver these bits to the row's PE. The up-scaled micro-architecture has 256 16x16 PEs and 256 HBMs to supply the weights and process the inputs in a single pass. To save resources, we propose 128 16x16 PEs and 128 HBMs and two passes to process all the inputs. We also use an HBM for the input memory so that 16 16 bit words can be read out in 1 cycle to provide the 16x1 input vector to each of the 16x16 PEs. The main control sequence in Fig.8 therefore reduces from 11 cycles to 7 cycles : 4 for reading the HBM with weights ( overlapped with 1 cycle for reading the HBM with inputs), and 3 for matrix-vector multiplication, accumulation, and write back.

Fig. 9 shows the scheduling with the up-scaled microarchitecture for FC6 and FC7 when 128 16x16 PEs and 128 HBMs are used with two passes. In the horizontal (time) direction, AlexNet FC6 requires 576 (9216/16) time slots per pass, VGG16 FC6 requires 1568 (25088/16) time slots per pass, and FC7 requires 256 (4096/16) time slots for either network.

Two pages in each HBM are required store all the weights for the HBM's row (PE). The first page is used in pass 1 and the second page in pass 2. The maximum number of weights for the FC6 layer in VGG16 is 25088x4096 or 102,760,448 weights. Using 2 bytes per weight, this requires 268 MB of storage which is easily stored in the new HBM2 16 GB part (Flashbolt) from Samsung. Each page stores 134 MB of weigths. The FC6 layer in AlexNet is 9216x4096 so it has less weights and can also fit in the 16 GB HBM2.

The switching between pages in an HBM is assumed to be negligible as well as the saving of the 128 PE's outputs after each pass to output memory. Using the largest layer, FC6 VGG16, each pass will require 1568 time slots; using a pipelined 16x16 PE at 662MHz and 7 cycles per time slot will therefore require 16.6 usec for a pass. Both passes will take 33.2 usec for processing the entire layer. This is compared to 34.4 usec as reported in Table IV in [12] which uses compression and is a considerable improvement since the saving is accumulated over each forward inference.

Table IV-D below summarizes the estimated total latency for FC6 and FC7 processing in both networks with the proposed up-scaled micro-architecture with pipelined 16x16 PEs. For comparison, latency from Table IV in [12] , which uses

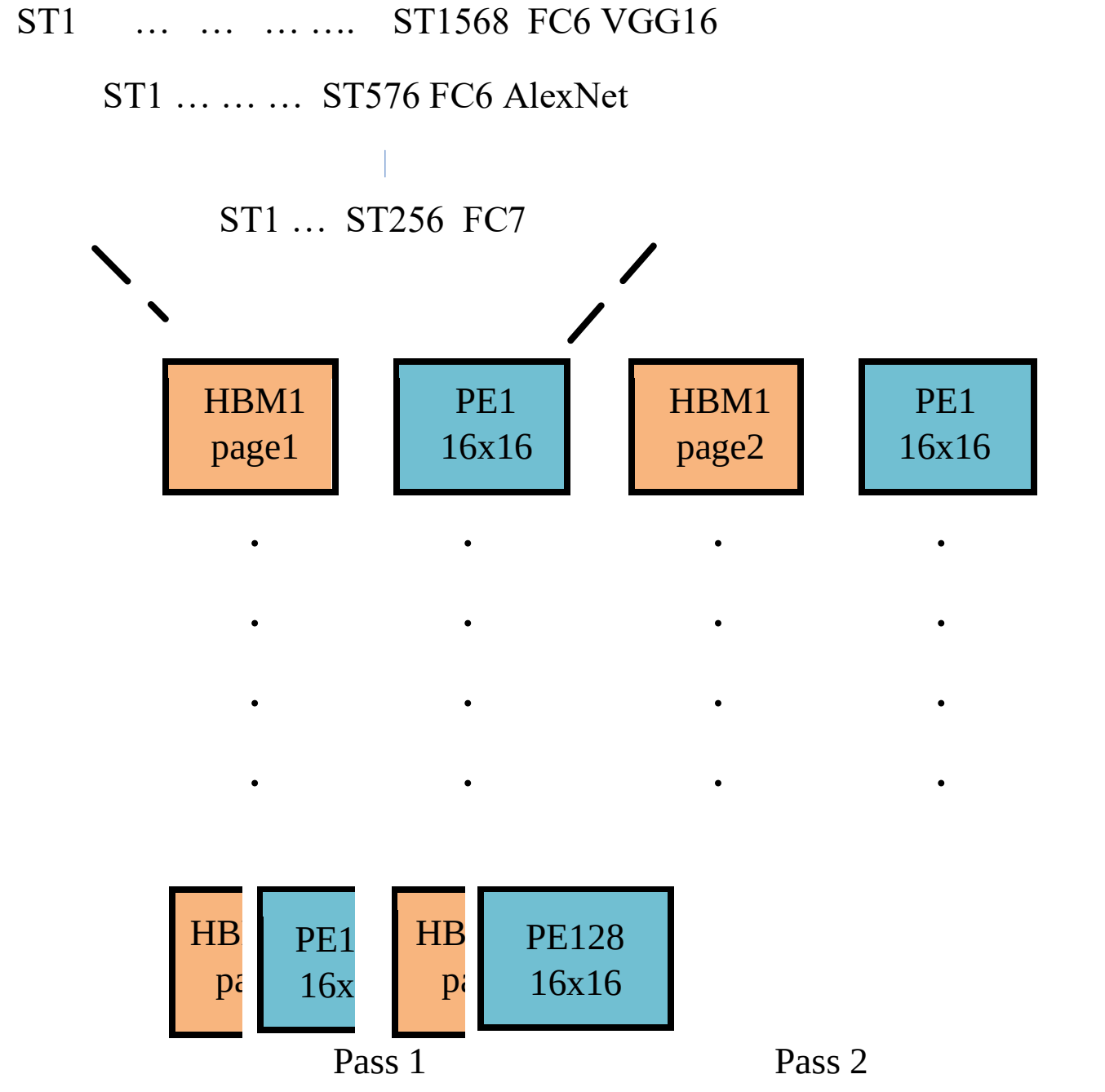

Fig. 10: Up-scaling the propsed micro-architecture for handling FC6 and FC7 layers. In each pass 128 HBMs and 128 16x16 PEs are re-used. The input and output memories are connected as in Fig.3

compression in each layer, is also included.

TABLE VI: Estimated Performance of FC6 and FC7 layers

| Layer parameter | AlexNet VGG16 | |
|---|---|---|
| this work FC6 latency | 12 usec | 33.2 usec |
| EIE [12] FC6 latency | 30.3 usec | 34.4 usec |
| this work FC7 latency | 5.41 usec | 5.41 usec |
| EIE [12] FC7 latency | 12.2 usec | 8.7 usec |

## V. CONCLUSION

We have discussed a novel HBM based micro-architecture for accelerating fully connected layers in DNNs and CNNs such as FC6,FC7, and FC8 in AlexNet and VGG16. For the FC8 4096-1000 layer in AlexNet and VGG16, we achieve 108 GOPS (non-pipelined 8x8 PE) with 100 MHz at 17 W, in PDK 45nm 1V, and 1048 GOPS (pipelined 8x8 PE) with 662 MHz in the same technology. Each PE is based on a 8x8 tile of weights. The achieved processing latency improves( 14 % reduction in FC8 AlexNet) on recently published results for the same FC8 layer without using compression. The microarchitecture can easily scale up to accelerate the larger FC6 and FC7 layers in AlexNet and in VGG16 with the same number (128) of paged HBM2E memories for the weights and 128 16x16 PEs to process larger 16x16 tiles of weights. Two pages in each HBM are required to support the larger layers. The estimated processing latencies for these layers is an improvement ( 60.4 % reduction in FC6 AlexNet and 3.49 % reduction in FC6 VGG16) , on recently published benchmarks for the EIE [12] accelerator of the same layers.

## REFERENCES

[1] V. Sze and et al, “Efficient processing of deep neural networks: A tutorial and survey,” *IEEE Proceedings*, vol. 105, no. 12, pp. 2295–2329, 2017.

[2] Q. Yuran and et al, “Fpga-accelerated deep convolutional neural networks for high throughput and energy efficiency,” *Concurrency Computat: Pract. Exper, Wiley Online Library*, vol. 29, no. e3850, pp. 1–20, 2017.

[3] Q. Jiantao and et al, “Going deeper with embedded fpga platform for convolutional neural network,” *ACM FPGA’16 Conference DOI: http://dx.doi.org/10.1145/2847263.2847265*, pp. 26–35, 2016.

[4] L. Ning and et al, “A multistage dataflow implementation of a deep convolutional neural network based on fpga for high-speed object recognition,” *IEEE Southwest Symposium on Image Analysis and Interpretation*, pp. 165–168, 2016.

[5] L. Huimin and et al, “A high performance fpga-based accelerator for large-scale convolutional neural networks,” *26th International Conference on Field Programmable Logic and Applications (FPL).*, pp. 1–9, 2016.

[6] NVIDIA, “cusparse,” *http://developer.nvidia.com/cusparse.*, 2018.

[7] JEDEC, “Jesd235a,b,c high bandwidth memory (hbm) 3d dram standard,” *www.jedec.org*, 2020.

[8] D. Hammerstrom, “A vlsi architecture for high-performance, low-cost, on-chip learning,” *Neural Networks, 1990., 1990 IJCNN International Joint Conference on*, pp. 537–544, 1990.

[9] T. Chen and et al, “Diannao: A smallfootprint high-throughput accelerator for ubiquitous machine-learning,” *ACM Sigplan Notices*, vol. 49, no. 4, pp. 269–284, 2014.

[10] Z. Du and et al, “Shidiannao: Shifting vision processing closer to the sensor,” *ACM SIGARCH Computer Architecture News*, vol. 43, no. 3, pp. 92–104, 2015.

[11] Google, “Edge tpu, google’s purpose-built asic designed to run inference at the edge.” *https://cloud.google.com/edge-tpu/*, 2020.

[12] S. Han and et al, “Eie: Efficient inference engine on compressed deep neural network,” *ACM/IEEE 43rd Annual International Symposium on Computer Architecture*, pp. 1–6, 2016.

[13] M. Gao and et al, “Tetris: Scalable and efficient neural network acceleration with 3d memory,” *Platform Lab, Stanford Univ. https://platformlab.stanford.edu/pdf/Mingyu<sub>G</sub>ao.pdf,pp.*1 – −25,2017.

[14] D. Kim, J. Kung, and et al, “Neurocube: A programmable digital neuromorphic architecture with high-density 3d memory,” *ACM/IEEE 43rd Annual International Symposium on Computer Architecture (ISCA)*, vol. DOI: 10.1109/ISCA.2016.41, no. 12, pp. 380–392, 2016.

[15] S. Li, “Towards efficient hardware acceleration of deep neural networks on fpga,” *Ph.D. Thesis University of Pittsburgh*, 2018.